# "Modeling somatic computation with non-neural bioelectric networks"

Santosh Manicka and Michael Levin*

Allen Discovery Center
200 College Ave.
Tufts University
Medford, MA 02155

* Author for correspondence:
    Email: michael.levin@tufts.edu
    Tel.: (617) 627-6161

# Abstract

The field of basal cognition seeks to understand how adaptive, context-specific behavior occurs in non-neural biological systems. Embryogenesis and regeneration require plasticity in many tissue types to achieve structural and functional goals in diverse circumstances. Thus, advances in both evolutionary cell biology and regenerative medicine require an understanding of how non-neural tissues could process information. Neurons evolved from ancient cell types that used bioelectric signaling to perform computation. However, it has not been shown whether or how non-neural bioelectric cell networks can support computation. We generalize connectionist methods to non-neural tissue architectures, showing that a minimal non-neural Bio-Electric Network (BEN) model that utilizes the general principles of bioelectricity (electrodiffusion and gating) can compute. We characterize BEN behaviors ranging from elementary logic gates to pattern detectors, using both fixed and transient inputs to recapitulate various biological scenarios. We characterize the mechanisms of such networks using dynamical-systems and information-theory tools, demonstrating that logic can manifest in bidirectional, continuous, and relatively slow bioelectrical systems, complementing conventional neural-centric architectures. Our results reveal a variety of non-neural decision-making processes as manifestations of general cellular biophysical mechanisms and suggest novel bioengineering approaches to construct functional tissues for regenerative medicine and synthetic biology as well as new machine learning architectures.

# 1. Introduction

Biological systems have long served as an inspiration and a design challenge for the engineering of artificial intelligence and machine learning[1,2], with a special focus on the brain[3]. However, many biological phenomena, ranging from maze solving by cells and slime molds to complex regulative morphogenesis and regeneration, can be viewed as processes involving information-processing and decision-making[4], in the absence of a brain[4-12]. Memory, anticipation, and problem solving have been demonstrated in sperm, amoebae, yeast and plants[8-12], and these capabilities scaled with the emergence of the Metazoa. For example, when craniofacial anatomical features of tadpoles are scrambled into abnormal configurations, they normalize their aberrant positions over time to regain a correct frog face morphology and then cease remodeling[13]. Thus, cells and tissues functionally ascertain the difference between the current, incorrect craniofacial morphology and the frog's native craniofacial morphology and undertake corrective movements to reduce the error. Similarly, the regeneration of entire limbs in animals such as salamanders can be understood as being driven by a 'test-operate-text-exist' model[7], in which cells act to implement an invariant anatomical outcome from diverse starting conditions, and stop once this target morphology has been achieved[7,14]. This theme is reinforced by classical observations such as the fact that tails grafted onto the side of salamanders slowly remodel into limbs[15], demonstrating the ability of tissue to ascertain its position within the whole, compare its organ-level anatomy with that dictated by the target morphology, and remodel toward that correct anatomical setpoint[16].

Importantly, unicellular life forms and somatic cells of multicellular organisms were making flexible decisions based on inputs in their microenvironment long before neurons appeared[11,17]. Nerves may have speed-optimized ancient bioelectric processes that, since the time of bacterial biofilms[18], were already exploited by evolution to implement memory, long-range coordination, and decision-making utilized for maintenance and construction of anatomical structures[19]. In multicellular organisms, these same functions are used to control large-scale patterning[20-23]. For example, bioelectric signals mediate important aspects of the long-range coordination that keep cells harnessed towards maintenance of a body-plan and away from tumorigenesis[5]. Stable bioelectric circuits also maintain the information needed for fragments of planaria to regenerate the correct number and distribution of heads and tails[24,25], and spatial distributions of resting potentials that are critical for the correct formation of hearts[26,27], eyes[28], and brains[29]. Bioelectric signaling is involved at many scales, from decisions of stem cells as to when and how to differentiate[30], to the control of size[31] and regeneration of entire organs[32,33]. Transient inputs that shift the endogenous bioelectric network states into alternate stable modes can permanently change regenerative and developmental morphology[34]; because of this, mutations in ion channel genes are an important cause of human birth defects[35,36], while modulation of bioelectric circuit state has been used to repair brain structure[37]. Thus, a number of important fields converge on the need to understand how tissues process information via bioelectric processes: evolutionary developmental biology (plasticity in cell behavior contributes to evolvability of body plans), regenerative medicine (induction of complex repair by modulating the patterns to which cells build), and synthetic bioengineering (the pursuit of novel synthetic 'living machines').

Decision-making often exploits some form of logic. The need to understand tissue functions as computation mediated by ancient biophysical mechanisms led us to seek a quantitative model of how non-neural networks could possess logical abilities like those commonly studied in brains. The simplest realizations of logic are known as "logic gates", which are tiny circuits that perform elementary logic operations like AND, OR, NOT, etc. that comprise any modern digital computer. A unique feature of logic gates is that although they are typically realized as electronic circuits, the underlying mathematical formulations have been used as coarse-grained models of neurons[38,39], genes[40,41], and even physiology[12]. Logic networks can be assembled to compute any computable function[42], and thus can be valuable building blocks of intelligent information-processing systems.

Here, we show that non-neural bioelectric networks can compute logic functions, suggesting one way in which evolution can exploit biophysics for decision-making in cellular systems. This provides a new connection between non-neural physiology and a common kind of computational task, thus expanding the known capabilities of developmental bioelectricity. While defining necessary and sufficient conditions for "computation" is a profound question debated elsewhere[43-46], logic gates are widely accepted as a non-controversial example of computation. Thus, we extend established connectionist approaches to a more general physiological setting and analyze them to reveal how ionic dynamics in non-neuronal cells could implement both simple and complex logic gates.

**A new bioelectric network model**

To facilitate the analysis of dynamic, context-dependent biological processes (e.g., morphogenesis and remodeling) as cognitive tasks[4], where networks of non-neural cells collectively make decisions, we constructed a minimal BioElectric Network (BEN) that comprises the simplest components and processes of bioelectrical signaling. BEN is inspired by a sophisticated and realistic model known as the *Bioelectric Tissue Simulation Engine*[47] (BETSE), which has been used to show how bioelectric patterns can be created and sustained[48-51] and how such a system might interact with genetic networks to give rise to morphological patterns[52]. Our goal was to define a generic, minimal biophysical system using realistic yet simple signaling dynamics, and study its computational properties in the absence of the specialized, highly-derived specifications that define neural networks. Thus, BEN is a simplified version of BETSE that is minimal enough to aid the investigation of the computational capabilities of a bioelectric system that have not yet been understood. The inputs and outputs of this network, representing the logic values, are represented by the bioelectric states of the cells (resting potential or $V_{mem}$).

BEN is defined by a set of bioelectric components and processes that dictate their production, transport and decay. BEN consists of two types of regions: an environment and a network of cells. A single cell in a BEN network consists of different types of proteins: ion channels and ion pumps. There are two type of molecules: ions (charged) and signaling molecules (generic uncharged molecules that may represent a ligand, a neurotransmitter, a secondary messenger etc.). BEN contains three types of ions: sodium ($Na^+$) and potassium ($K^+$), and chloride ($Cl^-$). Furthermore, there are two types of ion channels (IC) that each selectively allow the passage of $Na^+$ and $K^+$ ions respectively between the cell and the environment; no such channel exists for $Cl^-$. There is a single sodium-potassium ion pump that actively strives to maintain a non-zero membrane potential ($V_{mem}$). Two cells in a BEN network may be connected by an electrical synapse known as a "gap junction"[53] (GJ), represented as an undirected edge in the network, which allows the passage of molecules between the cells. The process that governs the passage of ions through either the GJ or IC is called "electrodiffusion", a combination of electrophoresis and regular diffusion, where the flux is driven by voltage gradients in the former and concentration gradients in the latter. Following the convention adopted in BETSE, we modeled electrodiffusion through GJs using the Nernst-Planck equation, and the transmembrane flux of ions through ion channels using the Goldman-Hodgkin-Katz flux equation[47]. However, since BEN is a networked system where the processes occur at discrete points in space (the cells in Fig. 1a), we used simple discretized versions of the same equations. The flux of the signaling molecules in BEN is governed by a generic nonlinear reaction-diffusion process consisting of two layers of sigmoidal transformations. In the first transformation layer, the flux across each GJ is calculated using a sigmoidal transformation of the concentration difference of the signaling molecule. In the second layer, the transformed flux across each GJ is collectively transformed using a second sigmoid, constituting the net concentration change of the signaling molecule. These two layers minimally represent the multi-layered complexity prevalent in cell signaling networks[54,55]. For example, calcium ($Ca^{2+}$) transport within a cell is mediated by clusters of inositol 1,4,5-triphosphate ($IP_3$) receptors distributed inside the cell in a multi-layered fashion[56-58]. Moreover, such a two-layer transformation is also found in neuronal cells, where the first stage involves nonlinear transformations of local dendritic potentials, and a second stage involving a linear summation of the outputs from the first stage to determine the overall cellular response[59-61]. Thus, the two-layer signal transformation architecture of BEN is representative of the multi-layered signaling complexity characteristic of multiple cell types. Moreover, the dynamics of the signal may be viewed as a reaction-diffusion system; see the 'Methods' section for more details. Finally, the extracellular environment simulated in BEN includes the same three types of ions as the cells ($Na^+$, $K^+$ and $Cl^-$), but initialized with different concentrations. We assume that the environment is effectively infinite in size, hence the ion concentrations there remain constant.

There are two types of gating mechanisms in BEN: chemical-gating of ion channels and voltage-gating of gap junctions. Chemical-gating of IC is a mechanism by which the permeability of an IC is modulated by the binding of a gating molecule (typically a 'ligand') to the channel, as a function of the concentration of the molecule[52,62(Ch.31)]. Rather than the more conventional Hill function to model it[52], we used a simpler sigmoid function that maps the concentration of the signaling molecule to the proportion of the maximum permeability of the IC. In BEN, a higher concentration of the signal tends to depolarize the cell, and lower concentrations hyperpolarize it. Voltage-gating of GJ is a mechanism by which the permeability of a GJ is modulated by the $V_{mem}$ of the connected cells and also the difference between them (known as the "transjunctional voltage")[52,63,64]. We adopted a sigmoidal transformation that maps the individual $V_{mem}$ of the connected cells that are then averaged over to compute the net GJ permeability, in contrast to the more complex but similar Hill function based model adopted in BETSE[52]. In BEN, positive $V_{mem}$ levels tend to make the GJ more permeable, and negative $V_{mem}$ make them less permeable. The mathematical details of BEN are described in the 'Methods' section.

There are two types of *learnable* parameters in BEN: weight and bias, following the convention of artificial neural networks[65] (ANN). The weight is associated with a GJ, and it helps modulate the flux of the signal as well as the voltage-gating of GJ. Negative weights tend to lower the signal concentration in a cell, and positive weights tend to increase it. With regards to voltage-gating, smaller weights diminish the effect of $V_{mem}$, and larger weights tend to magnify it. The bias is associated with a single cell that helps modulate signal flux: it specifies the threshold at which the signal concentration switches direction of change. By linking these parameters to the core features of BEN, we have assumed that they are not special to ANNs.

BEN networks have a layered architecture (Fig. 1a), a scheme that is widely adopted for ANNs[65] but which also reflects the tissue layering of many different types of body structures. In this scheme, a network consists of multiple layers of cells, with different choices for the inter-layer and intra-layer connectivity. Some layers have clearly designated roles—for example, the first layer is the "input" and the last layer is the "output". This follows the scheme of various biological systems with specific roles assigned to the layers. For example, the retina is the "sensory" layer that directly receives signals from the environment and sends it to the upstream inter-neuron layers that integrates the input[66]. The layered architecture is inspired by biological counterparts as diverse as the mammalian visual cortex[1,2,66] and cellular signaling networks[67], whose benefits include better organization and efficiency of information processing[65] and prediction[68].

A simulation of BEN proceeds as follows. The input layer cells are set to specific $V_{mem}$ levels at the beginning, while all other cells are set at $V_{mem} = 0$. From that point onward, the $V_{mem}$ levels of the network will dynamically change according to the network parameters, until the output settles at some $V_{mem}$. The input to the network is either *fixed* (clamped throughout a simulation) or *transient* (set at only the beginning of the simulation and then allowed to change). These two conditions represent the nature of biological information processing where the outcome may depend on the duration of the stimulus. For example, the induction of certain proteins in cellular signaling networks require long-duration input signals[69,70]; while a brief stimulus (progesterone) is sufficient to induce long-term regenerative response in *Xenopus* limb[71], a brief exposure to a gap-junction blocker is sufficient to cause stochastic phenotypic shifts in planaria[72], a transient invasion into a bacterial ecosystem can dramatically switch its character[73] and transient alterations to $V_{mem}$ cause permanent alterations to gene expression and phenotypic patterning in planaria[74].

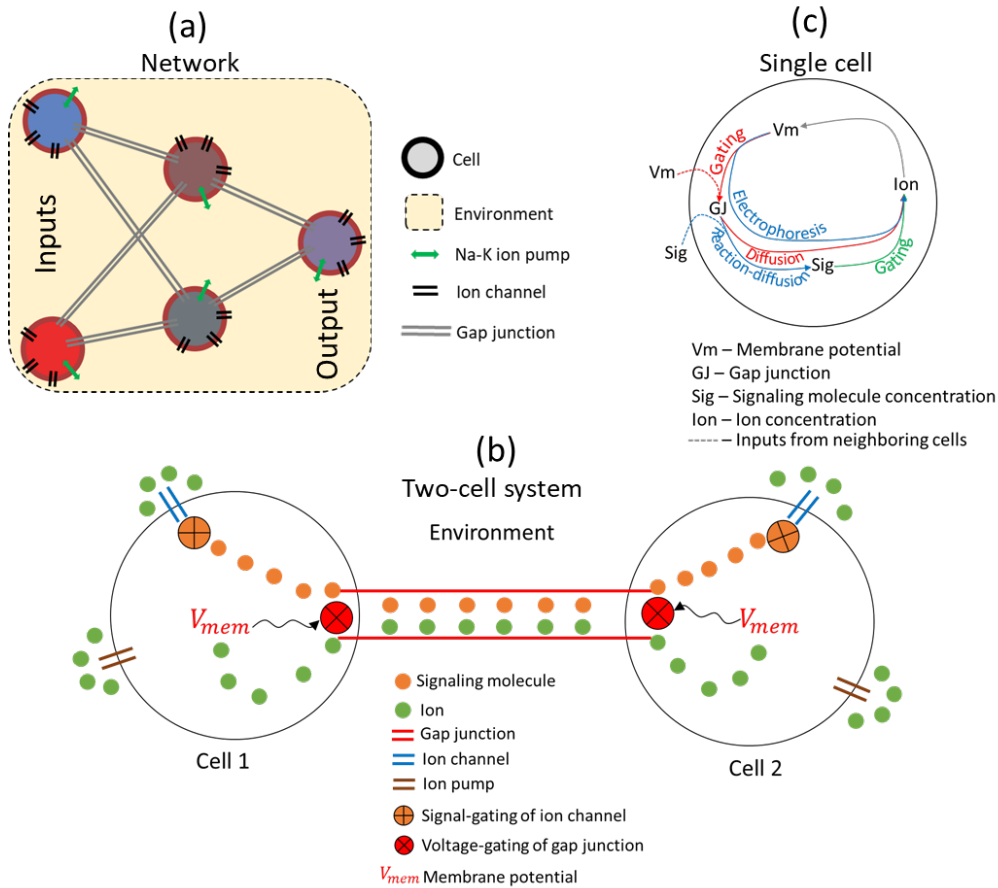

*Figure 1.* Schematics of the BEN architecture and its constituents. (a) The general architecture of a BEN network; (b) A BEN network consisting of two cells; (c) The network of components (black) related by processes (blue, red and green) within a single cell, some of which receive inputs from adjacent cells. For example, the $V_{mem}$ of a cell together with that of the adjacent cell gates the gap junction which in turn affects the concentrations of the ion and the signaling molecule, and the ion concentration ultimately determines the $V_{mem}$, closing the loop. This highlights the recurrent nature of the dynamics in BEN.

## 2. Results

The ability of somatic tissues outside the brain to compute is still controversial, in part because there is not a quantitative model available that demonstrates how basic operations of cognition could be implemented by generic cells. We demonstrate the logic-capabilities of BEN networks by constructing (1) small elementary logic gates; (2) larger "tissue-level" elementary logic gates; (3) compound logic gates composed from elementary gates; and (4) a pattern detector (showing how a complex regenerative response function can be implemented with these components) using standard machine learning methods. Below we show that BEN networks can indeed function as logic gates. We then analyze the behavior of a successfully trained BEN logic gate using the tools of information theory.

**BEN networks can implement small elementary logic gates**

A logic gate is a circuit that computes a binary-valued output from binary-valued inputs, according to a set of rules. For example, the AND gate outputs a "HIGH" signal only if both inputs are "HIGH", otherwise it outputs a "LOW". The OR gate, on the other hand, outputs a "HIGH" signal if at least one of the inputs has a "HIGH" signal. Thus, a logic gate is defined by a "rule table" that maps an input configuration to a unique output. Many biological systems appear to implement logic functions to facilitate, for example, integration of the various inputs they receive from the environment with the required specificity[55,75].

Here we demonstrate that BEN networks can function as logic gates. We used standard machine learning methods to identify the parameters of BEN networks that can perform a desired logical function. This discovery phase in the biological world can be implemented either by evolution (which tweaks the parameters on a phylogenetic time scale) or within the lifetime of a single animal by the dynamic adjustment of ion channel and gap junctional open states as a function of experience (plasticity)[76].

To provide a proof-of-principle that somatic bioelectric networks can perform logical operations, we sought to identify BEN networks with specific behaviors. Since it is difficult to find the requisite parameters manually, we used machine learning to automatically discover suitable parameter values. We used "backpropagation" (BP) for that purpose. BP is a training method, often used in machine learning, that starts with a random network (defined with a random set of parameters) and gradually tweaks the parameters using gradient descent until the network performs the desired function. Since BEN networks are recurrent by definition (edges are bidirectional), we used "backpropagation through time" (BPTT) to train them; see 'Methods' section for more details. We do not claim that BP or BPTT is the method that organisms use to learn in the real world; they are just convenient tools that we used to discover parameters that illustrate the power of BEN networks (see Supplementary 3 for alternative training methods). Below, we show examples of BEN networks implementing the AND gate, and the more difficult XOR gate. First, we show the results of training for a set of 100 training runs for each gate. Then, we describe the behavior of the best gate in each set using time series and dynamical phase space illustrations.

We first sought to determine if bioelectric networks can implement an AND gate, by training a small five-cell BEN network (Fig. 2a) to follow the AND rules that specifies the expected output for a given pair of inputs (Fig. 2b). We trained 100 instances of this network, each starting from different initial conditions of the parameters, discovering that some were indeed able to attain good performance, illustrating that BEN networks can perform the AND function (Fig. 2c). Out of the hundred networks we trained, seven achieved an error of 0.3mV or less, while a total of 33 networks achieved an error of 1mV or less. The behavior of the best AND gate we identified is illustrated below by way of time-series plots and phase space diagrams (Figs. 2d, 2e) by considering various input-output conditions.

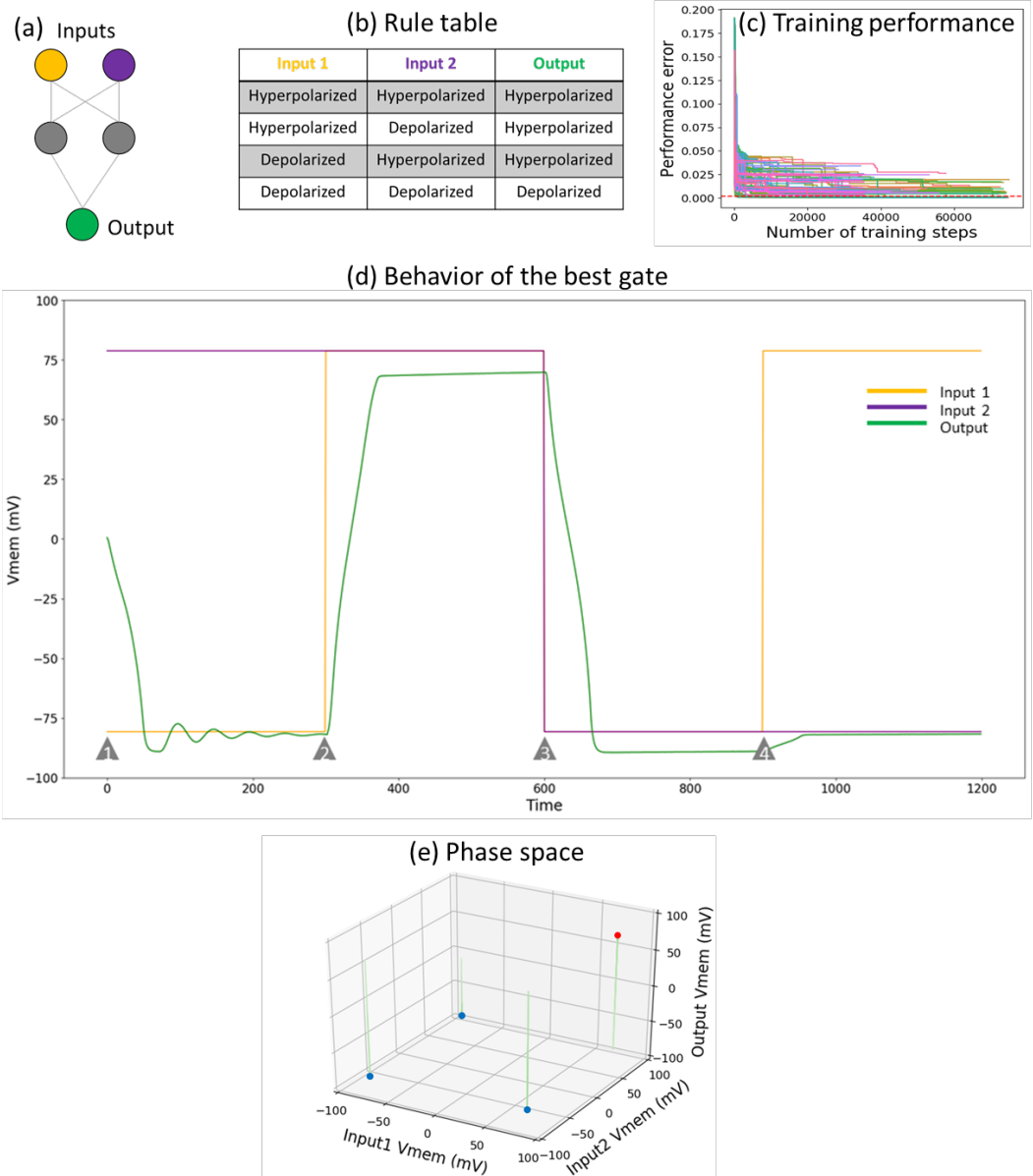

*Figure 2.* The AND gate. (a) A schematic of the AND logic gate and its rule table. The orange and purple cells are the inputs, and the green cell is the output; the grey cells are intermediate cells offering computational support. (b) Specification of the output $V_{mem}$ for various combinations of input $V_{mem}$ levels. This is known as the "truth table" in the Boolean logic literature, where the "hyperpolarized" state (around -80 mV) is indicated as "0", "False" or "OFF" and the "depolarized" state (around +80 mV) as "1", "True" or "ON". This rule table essentially summarizes the mechanism of the BEN-based AND gate: the output is depolarized only when both of its inputs are depolarized; in Boolean logic terms the output is ON only when both inputs are ON. (c) Pareto front of training errors over time (one unit is equal to a single training epoch) for the AND gate. This plot depicts the "front" with the best errors achieved over time. This Fig. demonstrates that the training does in deed result in learning. (d) The behavior of the best AND gate shown in the form of time series of the input and output nodes of the gate, shown for all four input-output conditions generated in a random sequence. The red and blue lines represent the states of the two input nodes, and green represents the output. The grey triangles mark the time points at which the inputs are switched to a different state. For example, both inputs are hyperpolarized at time point (1), while at time point (2) the red input is depolarized. (e) The dynamical phase of the logic gate: a depiction of a set of trajectories in the input-output space, illustrated in a time-lapse style. This dynamical system has two attractors in the output space,

highlighted in filled red (depolarized state) and blue (hyperpolarized state) circles. The trajectories look straight because the inputs are fixed, and only the output changes.

We next discovered a small five-cell BEN network that can function as an XOR gate by following the same training procedure as for the AND gate above (we report the parameter details in Supplementary 1). The rule table, training performances, behavior of the best XOR gate and the phase space diagram are shown below (Fig. 3). Of the fifty networks we trained, only one network achieved an error of 1 mV. Compared to the AND gate, clearly training the XOR gate is more difficult. This is consistent with the known difficulty of training any dynamical network to behave as an XOR function[65], since it is a nonlinear function compared to AND, which is linear.

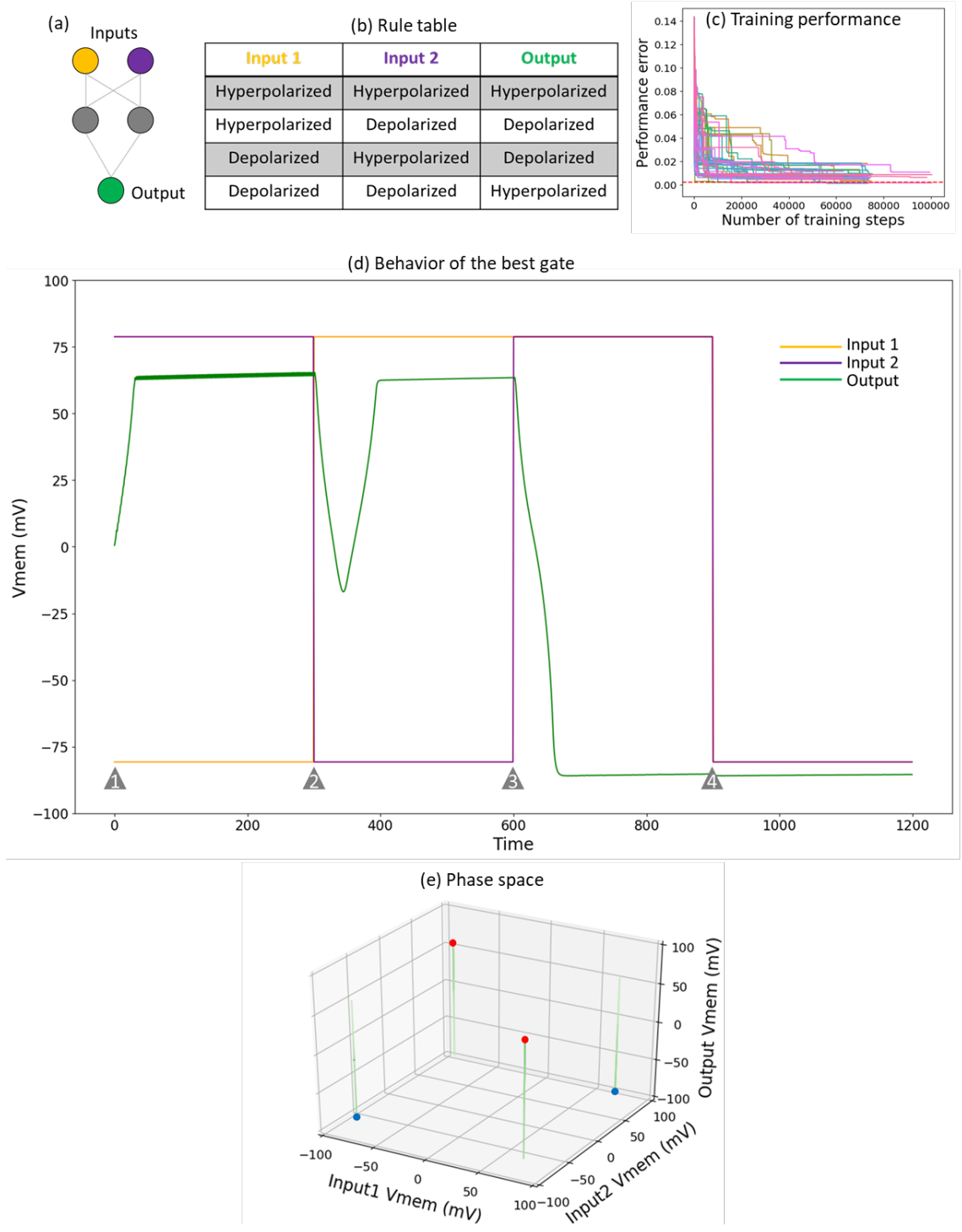

*Figure 3.* The XOR gate. (a) Schematic of the XOR logic gate; the orange and purple cells are the inputs, and the green cell is the output; the grey cells are intermediate cells offering computational support. (b) The XOR rule table: output is depolarized only when one of its inputs, but not both, is depolarized. Table specifies the output $V_{mem}$ for various combinations of input $V_{mem}$ levels; this is known as the "truth table" in the Boolean logic literature, where the "hyperpolarized" state (around -80 mV) is indicated as "0", "False" or "OFF" and the "depolarized" state (around +80 mV) as "1", "True" or "ON". (c) Pareto front of training errors over time (one unit is equal to a single training epoch) for the XOR gate. This plot depicts the "front" with the best errors achieved over time. This Fig. demonstrates that the training does in deed result in learning, but fewer networks are successful compared to AND (Fig. 2). (d) Behavior of the best XOR gate. Shown here are the time series of the input and output nodes of the gate, shown for all four input-output conditions generated in a random sequence. The red and blue lines represent the states of the two input nodes,

and green represents the output. The grey triangles mark the time points at which the inputs are switched to a different state. (e) The dynamical phase space of the gate: a depiction of a set of trajectories in the input-output space, illustrated in a time-lapse style. This dynamical system has two attractors in the output space, highlighted in filled red (depolarized state) and blue (hyperpolarized state) circles. The trajectories look straight because the inputs are fixed, and only the output changes.

**BEN networks can implement larger "tissue-level" elementary logic gates**

We next asked whether larger networks, which contain more cells than strictly necessary, could still perform the logic functions. These larger networks are like tissues and organs, which can contain a range of diverse cells that communicate. Thus, we next describe a "tissue-level" AND gate. First, we show the results of training for a set of 100 training runs for each gate. Then, we describe the behavior of the best gate in each set, using time series and dynamical phase space illustrations.

We used a combination of genetic algorithm (GA) and BP for identifying this kind of circuit; the GA discovers the network structure (example structure in Fig. 4 inset) and BP finds the parameters of the network we sought. The best network was obtained at the end of 233 generations, which achieved an average error of about 0.6mV. More details of the training are described in the 'Methods' section.

We conclude from the above results that it is indeed possible for large BEN circuits to compute simple logic functions, suggesting that this type of signaling can be carried out by developmental compartments with varying cell numbers.

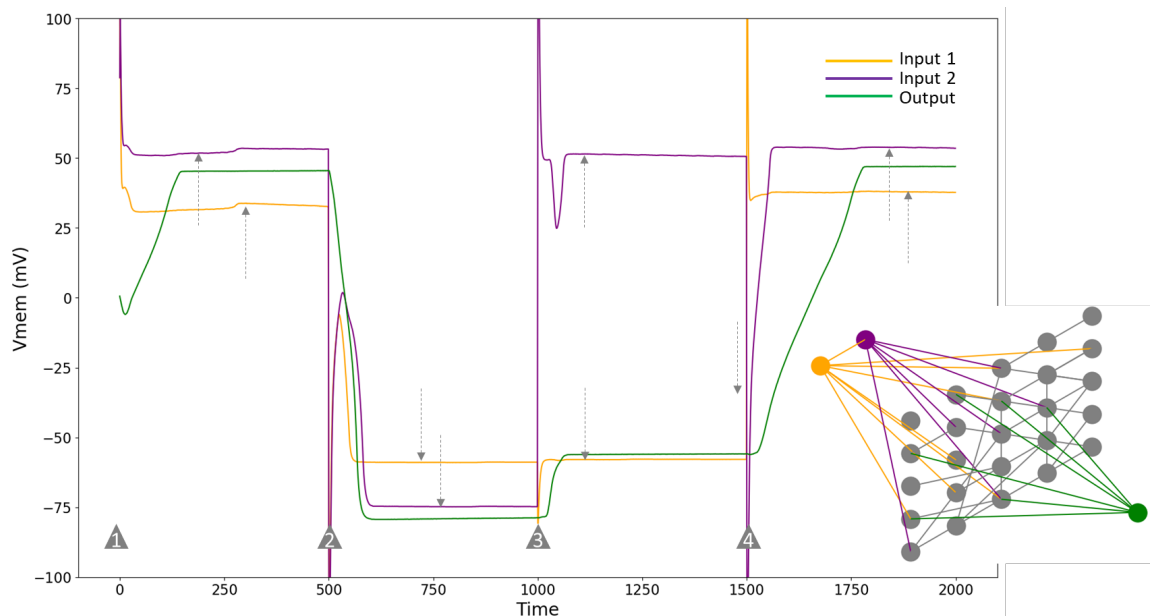

*Figure 4.* The structure and behavior of the best evolved tissue-level AND gate. Following the convention of the color-coding, orange and purple represent input nodes, while green represents the output node. Behavior correctly follows the AND rule (Fig. 2b). Inputs are transient: they are set (externally) to their respective states at the time steps marked by the grey triangles and then removed. As can be seen, they slightly oscillate for a small period of time after they are set initially, before dynamically fixing themselves (marked by grey dashed arrows) during every simulation. Thus, the network has evolved the capacity for memory: it remembers the inputs (at least their qualitative levels) even after they are removed. Inset shows the structure of the network.

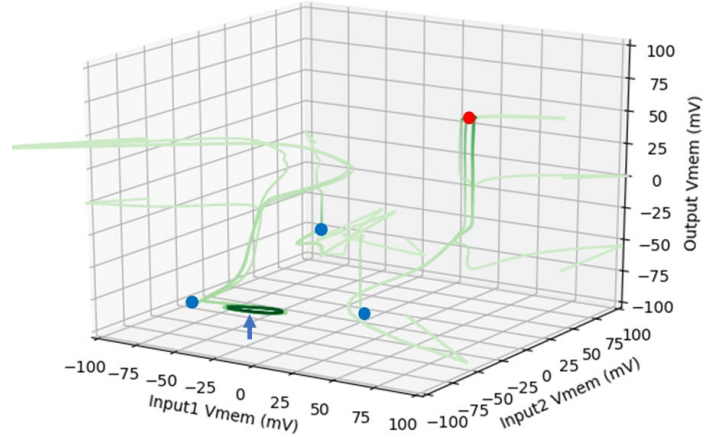

*Figure 5.* The dynamical phase space of the *best* tissue-level AND gate. A depiction of a set of trajectories in the input-output space, illustrated in a time-lapse style. This dynamical system has two attractors in the output space, highlighted in filled blue (hyperpolarized state) and red (depolarized state) circles. As can be seen, there are three standard hyperpolarized attractors (fixed-point-like), and a fourth cyclic hyperpolarized attractor (marked with a blue arrow). Even though the cyclic attractor is in the appropriate region of the phase space, it was not actually required for during the training—it accidentally emerged. Furthermore, as opposed to the straight trajectories of the previous phase space diagrams, the trajectories maneuver freely here, and occasionally spill over the usual $V_{mem}$ range which is about [-100,100] mV. This is because the inputs are transient and not fixed, hence they change states as well over time.

**BEN can implement a complex "tissue-level" logic gate: a pattern detector**

One of the most impressive computational tasks that tissues undertake is the set of decisions that enable large-scale regeneration of organs and appendages in some species[19,77]. Specifically, to repair after (unanticipated) injury and stop growth and remodeling when the anatomy has been corrected, cells need to make decisions about the physiological and geometric state of other tissues. Cells need to ascertain whether a large-scale morphology is correct or not, in order to regulate regenerative pathways or cease further change (errors in achieving morphostasis can manifest as cancer). Much work has gone into characterizing the physiological events that signal the binary event of "injury", but it is now clear that even in the absence of trauma, some complex organ systems such as the craniofacial structures[78,79] can begin drastic remodeling when the configuration is incorrect. Despite the advances in molecular biology that have identified genes *necessary* for this to occur, it is entirely unclear how cells recognize correct vs. incorrect patterns on a spatial scale much larger than themselves. Thus, we next sought to demonstrate that realistic biophysical mechanisms can implement computations which are *sufficient* to enable this crucial capability of living systems.

A pattern detector can be thought of as a complex (more than two inputs, thus non-elementary) AND gate, since its output is ON only when each of the inputs is at a specific desired state—in other words, only when *all* input conditions are satisfied. This is a more realistic setting that the tissue-level AND described in the previous section, since it has many more than two inputs (a tissue typically consists of many more than two cells). We identified, as proof-of-principle, a relatively large (44 cells) non-elementary logic gate that recognizes French-flag-like patterns; we used the French-flag as the pattern of interest due to its prominence in developmental biology as a simple morphogen gradient[80]. We define a French-flag pattern as a particular configuration of $V_{mem}$ levels in the input layer (3x6): the leftmost two columns (band) are blue (hyperpolarized $V_{mem}$), the middle band is grey (intermediate $V_{mem}$), and the rightmost band is red (depolarized $V_{mem}$). This network recognizes the French-flag pattern and slightly noisy versions of it by expressing a depolarized $V_{mem}$ in the output cell, while expressing a hyperpolarized $V_{mem}$ in the output as a response to all other patterns.

The pattern detector is set up similar to the tissue-level AND gate. The main differences between the two are: (1) inputs are fixed in the pattern detector, as was the case for the small logic gates described in Section 4.1; (2) inputs are set after an initial time period during which the network is allowed to settle at its intrinsic "baseline" state (this captures the essence of biological systems that have intrinsic activity even in the absence of external stimuli); and (3) input patterns are generated randomly where two sets are generated, one consisting of the French-flag and noisy variants of it, and the other consisting fully randomized versions of the French-flag (details in the 'Methods' section). We used the combined GA-BP search-train method, as before, to find this network; see more details in the 'Methods' section. The best evolved pattern detector is the individual with the highest fitness score in the last generation of the GA, which was discovered relatively quickly at the end of 11 generations (three other networks achieved similar errors). Even though the pattern detector is larger than the tissue-level AND described above, it was easier to train than the latter. This is due to the advantage of the inputs being fixed in the case of the pattern detector but transient in the case of the tissue-level AND gate: the fixed inputs provide a constant supply of information that the network does not have to remember unlike the tissue-level AND gate.

The above search successfully discovered a successful French-flag detector. The behavior of this detector when it sees a French-flag pattern and a random pattern in its input layer is depicted in Fig. 6a: in the first case the output is depolarized (red), while in the second case it is hyperpolarized (blue), both as expected. Furthermore, this detector recognizes slightly distorted variants of the French-flag where the distance of a pattern from the French-flag is the Euclidean distance (Fig. 6b). We observe that input patterns up to a distance of about 150mV from the French-flag are recognized as French-flag, while those at a distance of about 350mV and above are classed as not French-flag. Finally, the detector also responds correctly to a *sequence* of randomly chosen input patterns (Fig. 6c), showing that the behavior of the network only depends on the state of the inputs and not on the rest of the network.

We conclude that BEN has in principle the ability to distinguish between patterns. The biological implication is that somatic tissues can in principle have the same ability, and thus be a part of much larger pattern regulation mechanisms.

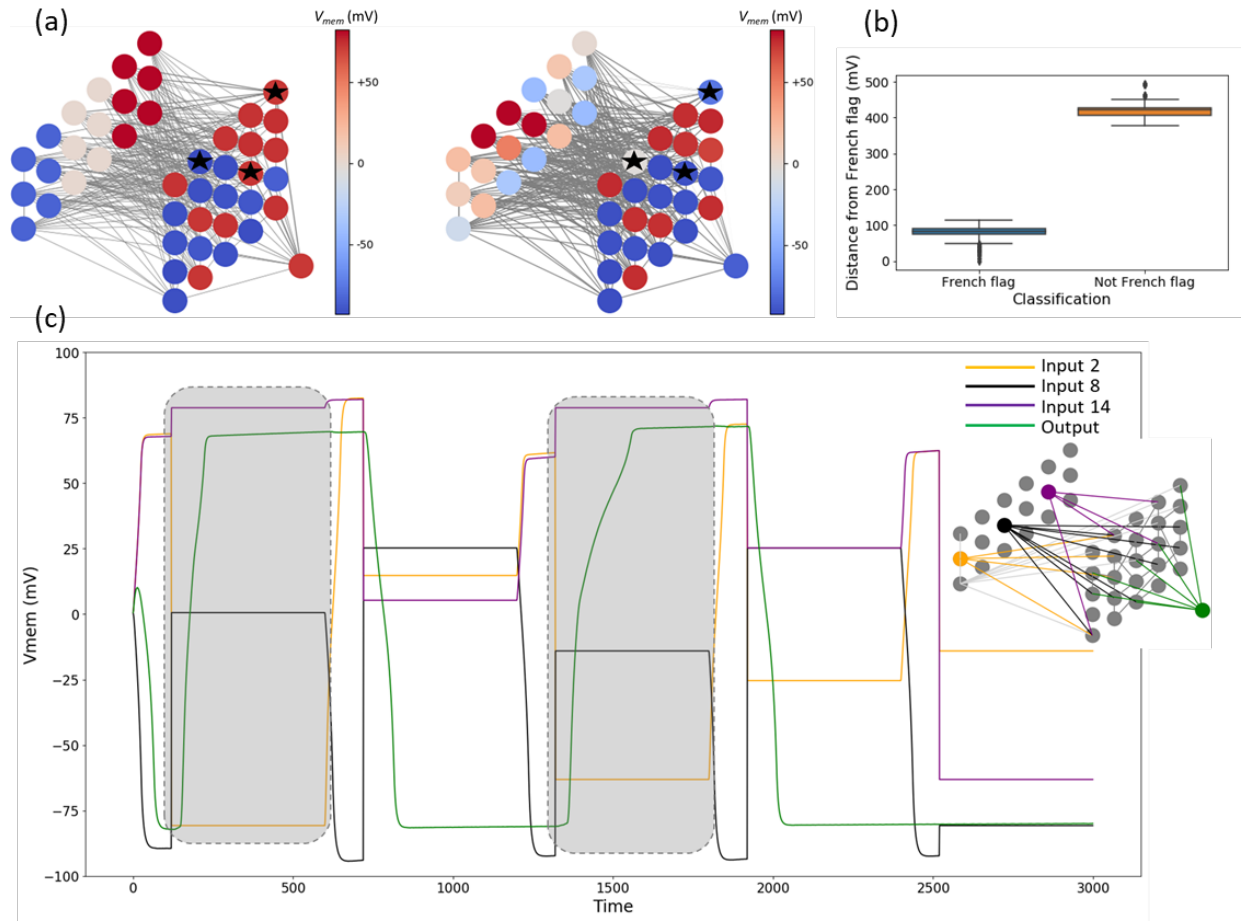

*Figure 6.* The French-flag detector. (a) Snapshots of the equilibrium (final) network states ($V_{mem}$ levels) in two detection scenarios. Left: input pattern is the French flag, and the output is depolarized (representing "correct" pattern). Right: input pattern is a highly distorted version of the French flag, and the output is hyperpolarized (representing "incorrect" pattern). Colors represent polarity levels of the cells in units of mV. Notice that the only three intermediate cells that differ in their states between the two cases are those marked with a black star, suggesting that the detector makes minimal use of information to distinguish between patterns. Also notice that the edges are thicker in the right than in the left, suggesting that the voltage-gating dynamics of the edges (representing gap junctions) play a crucial role (b) The overall performance of the best evolved pattern detector, with data collected from a set of 1000 simulations: 100 parallel sets each with a random sequence of 10 simulations. As expected, the pattern detector classifies patterns similar to the French flag (a total of 500 sample inputs) as "French flag", and those that are dissimilar (500 inputs) as "not French flag". This suggests that the pattern detector is robust to noise to a sensible extent. The width of the classification boundary was set implicitly due to the way sample input patterns from the two classes were generated (details in the 'Methods' section). (c) The behavior of the best French-flag detector shown for a set of four representative cells (three inputs out of a total of eighteen and one output) for a random sequence of five patterns. Inset highlights the four nodes whose colors correspond with those in the time series. Highlighted in grey squares are the cases where a French-flag-like pattern is input for which the output is depolarized, as expected; for all other cases where a random pattern is shown, the output is hyperpolarized.

**BEN can implement compound logic gates**

Biological processes are complex by nature. The rampant complexity is partly managed by nature by way of modularity[81], in gene regulation[82] and the brain[83,84], for example, at multiple scales. Modularity is beneficial because the modules can be independently tinkered with for the purpose of large-scale outcomes[82]. Thus, it is important to understand whether bioelectric circuits can implement logic functions that are too complex to be described by a single gate but are actually combinations of elementary logic primitives.

Compound logic gates are compositions of modules of logic gates, and their biological equivalents underlie several processes including sporulation in *B. subtilis* and the neuronal dynamics of *C. elegans*[85,86]. They can also have pharmaceutical applications—for example, the control of bacterial invasion of tumor cells and diagnosis of cellular environments and automatic release of drugs[85].

One way to design compound networks, in general, is to compose them from appropriate pre-designed modules. For example, a "NAND" gate can be constructed by combining the AND and NOT gates. Compound genetic circuits have been constructed in this fashion[87,88], as well as designed *de novo* from scratch[85]. Here, we show how to construct compound logic gates in physiological networks by composing pre-trained BEN logic gate modules. Unlike genetic networks and neural networks, which are directed, BEN networks are bidirectional, due to the symmetric nature of gap junctions. Thus, unlike genetic networks, compound logic gates in BEN cannot be constructed simply by connecting the modules. Put more precisely, if the modules were simply connected together, then the output of one module will receive signals from the input of the connected module. This could potentially result in the upstream module outputting the wrong state, and thus the downstream module receiving and outputting the wrong states as well. To mitigate this, we included a "bridge" layer that interfaces the modules and trained it so that the compound gate as a whole behaves as desired: the downstream module output the correct state for every set of inputs received by the upstream module. For more details on the architecture and training of the bridge, see Supplementary 4. Fig. 7 shows an example of a successfully trained NAND gate by composing pre-trained AND and NOT gates.

We conclude that it's possible to construct compound logic gates by composing pre-trained modules and training only an interface bridge connecting the modules. This method can in principle be extended to more complex circuits.

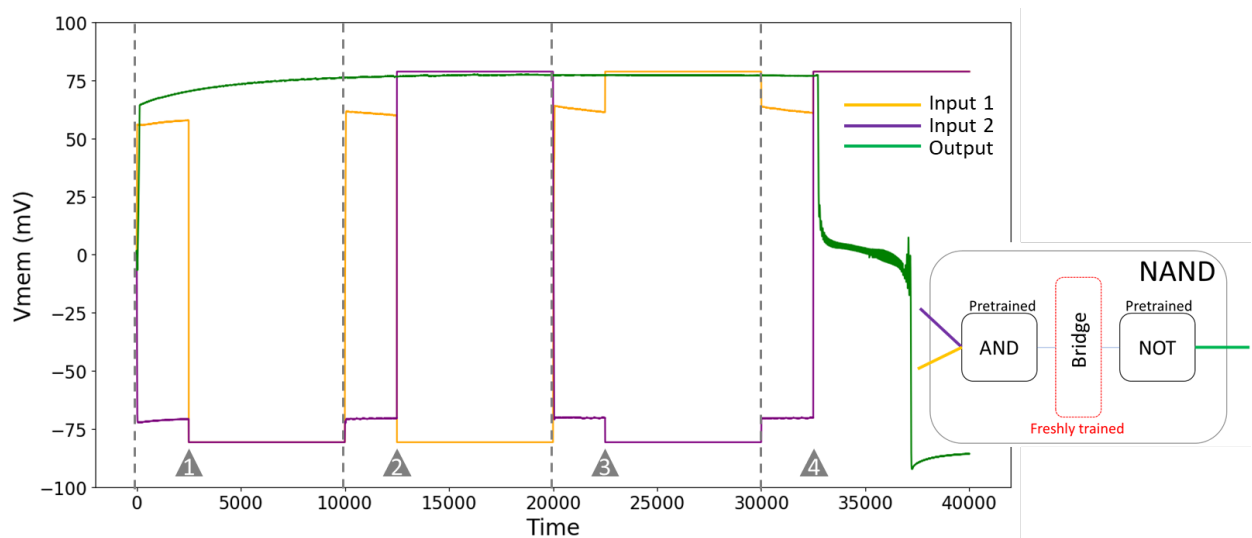

*Figure 7*. The behavior of a successfully trained NAND gate for a random sequence of four inputs. Orange and purple lines represent the activities of the inputs, while green represents the output. The vertical dashed lines mark the beginning of each trial in the sequence, however the inputs are applied, after an initial transient, at the time points marked by the grey triangles. Inset is a schematic of the architecture of the gate: pretrained AND and NOT gates are connected by a "bridge", a single layer of cells, that is alone trained.

## 3. Discussion

Various forms of natural and artificial computing systems with unconventional mechanisms have been proposed[89]. Even the modern von Neumann computer architecture was inspired by the mechanisms of DNA transcription and translation[90]. Other examples include "liquid computers" that involve interactions of fluid streams and droplet flows to perform logic operations[91]; "chemical computers" that utilize the dynamics and equilibrium-behavior of chemical reactions[92-96] for logic operations and learning[97,98]; DNA-based computers that utilize various forms of catalytic reactions such as 'seesawing' and 'strand-displacement' for pattern recognition[99], protein-DNA interactions to implement logic functions[100] and even for general-purpose computing[101]; gene regulatory networks that take advantage of the multistability of gene activity for learning[102], memory[103] and logic[87,88,104,105]; "bacterial computers" that use enzyme-based computing to implement logic gates[106]; electric current flow making use of Kirchhoff's law for solving mazes[107]; and even "sandstone computers" that utilize natural erosion processes to implement logic gates[108]. BEN is an unconventional computing system, since it's a non-neural system like the examples above.

Our work does not require a commitment to the view that every biological question must be dealt with via a computational perspective, despite the fact that there is a robust body of new experimental work driven by the hypothesis that some tissues are indeed implementing processes we would recognize as memory, comparison, or measurement[54,109-111]. The key question we answer here is: are pre-neural physiological networks able, in principle, to perform simple computations such as logical operations? Prior to these results, it was not known, and many developmental biologists' intuitions are that such networks cannot perform such functions - they are almost never considered for model-building. Our data provide the first proof-of-existence showing that the class of biological systems where a computational approach *could* be useful, such as evolved or synthetic regulative morphogenesis, needs to be expanded to include somatic bioelectric networks.

BEN is also a type of patterning system. Patterning is the ability of a system to generate, sustain and recognize patterns in the states, whether physical, chemical or biological, of the system. It is a widespread phenomenon reported in biological systems, including pattern recognition of pathogen molecules by cellular receptors[112], axial patterning in planaria[113], spatial patterning during neural patterning[114], cell polarity[115], genetic patterning in bacterial populations[116] and voltage patterning in neural tubes[50]. Here we focus on bioelectric non-neural networks of patterning. Various non-neural network models of patterning have been proposed. Some examples include the "packet-routing" system that uses an internet-like intercellular messaging system[117], auto-associative dynamical systems[118], and cellular automata[119] based systems that combine local and global communication to model pattern regeneration. Our laboratory has produced sophisticated non-neural bioelectric network models of tissue patterning, like the *Bioelectric Tissue Simulation Engine* (BETSE) that incorporates detailed biophysical mechanisms of ion channels and transport processes[47,52,120], as well as relatively simpler models that incorporate only the high-level functionality of the bioelectric components[49,121,122]. These models have been used to show that networks of ordinary tissues can give rise to gradient-patterns[47,52], Turing patterns[48,122], as well as oscillatory patterns[121,123] that are thought to be essential for long-distance communication. BEN differs from these models on two fronts: (1) it is biophysically simpler than BETSE, serving as a minimal model of dynamics sufficient for computation; and (2) it is more realistic than the model of Reference [121] which uses "equivalent circuits" where the equilibrium $V_{mem}$ is explicitly specified in the equations, whereas in BEN the equilibrium $V_{mem}$ emerges from the ion channel parameters.

We have shown here that BEN networks can implement elementary logic gates, more complex tissue-level logic gates, pattern detectors and compound logic gates. Furthermore, we have demonstrated that logic can be implemented in circuits with bidirectional connections that is typical of non-neural tissues, in contrast to the conventional directed circuits like neural networks and digital electronic circuits. This implies that even though non-neural tissues may find it harder to implement formal logic (due to lower control over information flow; see Supplementary Fig. S2b, for example, where information flow is recurrent even in a small network, making it hard to control), they can achieve it. Not only can BEN networks compute, but they can also be robust to damage. One of the most important, and heretofore unexplained, properties of biological control circuits is that they continue to function computationally under significant perturbation (as seen by work on metamorphosis, stability of memory and behavioral programs during drastic tissue remodeling, and invariance of morphogenesis to changes in cell number[79,124,125]). BEN networks can exhibit a similar behavior, retaining function even after the removal of cells. We found that BEN achieves this by distributing information in the network (Supplementary 2).

The particular methods by which BEN networks were trained is not of central importance in this work and do not impact the main result – that BENs can be readily found which implement logic functions. We used BPTT simply as a way to discover the parameters of (train) BEN networks that perform specific logic functions or recognize patterns. In the biological world, training (learning) may happen in other ways that may be broadly classified as offline and online. Offline learning happens outside of the lifetime of an individual and at a population level, while online learning happens during the lifetime of an individual. Evolution is an example of the former, while Hebbian learning is an example of the latter. BPTT may be classified as a form of online learning, but it has been claimed to be not biologically realistic[126,127]. More biologically realistic forms of online learning have been proposed, for example, forward propagation of eligibility traces[126], extended Hebbian learning[128], dynamic self-organizing maps[129] and instantaneously trained neural networks[130]. Such methods could potentially advance biologically realistic training of BEN networks, which we will develop in future research, but they are not necessary to find examples demonstrating that BEN networks can compute.

BEN models the bioelectric *principles* of a generic bioelectric network, and not *all* the physiological details, as these can vary widely across tissues. Individual cell types can be modeled in future effort simply by altering the ion channel parameters while retaining the basic form of the equations. Although BEN is a model of a non-neural bioelectric network, it has certain features that resemble those of a neural network, as described above (the weight and bias parameters, and the two-level nonlinear signal transformation). This may suggest that neural-like features are necessary for computation even in a non-neural network. On the other hand, it also suggests that non-neural features may support neural computation. Indeed, this is the subject matter of ongoing debates in neuroscience about how much glia contribute to neural computation[131,132].

Our work contributes to the field of basal cognition by describing how a simple network of aneural cells may perform both basic and complex logic operations. However, our goal is not to validate basal cognition as a field – it is just one context (besides others, such as developmental neurobiology, evolutionary cognitive science, and synthetic bioengineering) which our findings impact. Even if some computational views of specific biological systems turn out to be wrong, our results provide value in showing how physiological circuits can be made with specific (and useful) behaviors, both for synthetic biology applications and to understand functional aspects of developmental physiology, the evolution of brain circuits, etc. In prior work, a computational perspective of cell behavior [14,16,133] allowed our lab and other groups to formulate (1) ways of experimentally controlling large-scale anatomical outcomes [78], and (2) formulate and solve the "inverse problem"[14] of how the cells might actually achieve specific anatomical configurations, given cellular constraints and a range of starting conditions [79,134]. Our current work addresses (2) by demonstrating a possible approach that cells could use to implement logic functions under realistic cellular constraints.

Overall, BEN has the potential to shed light on as-yet-unexplained non-neural bioelectric information-processing. For example, when certain planarian species (highly regenerative animals) are cut they sometimes develop two heads, instead of a head and a tail, as a response to a temporary exposure to 1-octanol, a gap junction blocker. Without the gap junction blocker, they always develop a wild type morphology consisting of a single head and a tail. It has been shown that this decision to switch morphologies is bioelectrically controlled and non-neural in nature[72,135]. We hypothesize that the planarian bioelectric system encodes a pattern that either consists of a record of the gap junction blocker or not, and then maps it into different morphological decisions (wild-type or double-head) in a systematic way. We currently have a prototype BEN model that reproduces this phenomenon in a qualitative way, which we plan to report in the future. Finally, we have demonstrated in this work that the slower (reason described in 'Methods'), continuous mode of non-neural bioelectric signaling, compared to the faster, pulsating mode of neural signaling, is sufficient to perform logical computations usually associated with the brain. This firmly supports the possibility of somatic computation in real biological systems. How it compares to the timescales of genetic signaling and how it could facilitate bioelectricity to be upstream of genetics[54,106] is an open question that our future investigations will answer.

Our work paves the foundation for future research in training actual biological tissues for somatic computation. Ongoing research in our laboratory has identified ways by which gap junctions can be manipulated and controlled, further opening up the possibilities on this front. Overall, our research provides the conceptual and modeling foundations to understand and manipulate development and regeneration, and to construct computational synthetic living machines.

## 4. Methods

**Mathematical details of BEN**

The equations that define a BEN model are shown in Fig. 8 using an example 2-cell system. The equations are shown for cell-2 only; the same equations with appropriate change in subscripts apply to cell-1 and likewise to any other cell in larger networks. The processes that the equations represent, like electrodiffusion, reaction-diffusion and gating, are marked appropriately in the figure, thus self-explanatory. The following remarks serve to better illustrate the model. As described in the introductory section, electrodiffusion utilizes two types of gradients, namely voltage and concentration, to drive the flux. Those gradients are represented by the terms $v_1(t) - v_2(t)$ and $c_1(t) - c_2(t)$ respectively in the electrodiffusion equations. The ion-pump equations are approximate versions of the full mechanism modeled in equations (23-29) of BETSE[52]. BEN makes use of just the terms involving the ion concentrations in equation (27) of BETSE[52], as it lacks the other biochemical components like ATP, and by choosing appropriate term coefficients to compensate for the approximation it effectively preserves the relationship between ion concentrations and the pump flux.

The dynamics of the signaling molecule is viewed as a nonlinear reaction-diffusion process as it can be abstractly represented as: $R(D(\nabla c))$, where $\nabla c$ is the concentration gradient, $D(.)$ represents the 1st layer of sigmoid approximating a nonlinear diffusion, and $R(.)$ represents the 2nd layer of sigmoid approximating a nonlinear reaction. Moreover, the nested form of $R(D(.))$ departs from the conventional form of reaction-diffusion which is $\frac{dc}{dt} = D(c) + R(c)$, for example, $\frac{dc}{dt} = \nabla c + c^2$. Nevertheless, it qualified as reaction-diffusion since it clearly involves diffusion ($\nabla c$) and a nonlinear component (the sigmoid) that alters the net mass of $c$, representing production or decay depending on the sign of $\frac{dc}{dt}$. Another important point to note is that even though sigmoid functions, which define the signal flux, also constitute the activation functions of neurons in neural network models[38,65], there is a crucial difference: while neuron models are input-activated since neuronal synaptic junctions are directional, a cell in BEN is gradient-activated since gap-junctions are bidirectional. This is also partly the reason why non-neuronal cells tend to be slower than neural cells—diffusion is a slower process than directed currents.

There's no set range of $V_{mem}$ levels that a given BEN network will not exceed (as there are no explicit bounds on the ion concentration levels), but there are typically set to lie within [-80 mV, +80 mV]; see for example Figs. 2d, 3d, 4, 6 and 7. The concentration of the signaling molecule, on the other hand, is forced to lie with [0,1], that is, within the generic minimum and maximum possible values of the actual concentration which is not explicitly considered in BEN.

The software to simulate, train and evolve BEN networks is available at: https://gitlab.com/smanicka/BEN

**Ion flux with respect to cell 2 (electrodiffusion):**

Gap-junction flux $\{\phi_g^i = D_g(t)\left(c_1^i(t) - c_2^i(t)\right) + \frac{z^i q}{k_b T} D_g(t) c_2^i(t)(v_1(t) - v_2(t))$,
$\forall i \in \{Na, K, Cl\}$

Ion-channel flux $\{\phi_c^i = \frac{z^i v_2(t) F}{RT} D_c^i(t) \left( \frac{c_2^i(t) - c_e^i e^{\left(-\frac{z^i v_2(t) F}{RT}\right)}}{1 - e^{\left(-\frac{z^i v_2(t) F}{RT}\right)}} \right), \forall i \in \{Na, K\}$

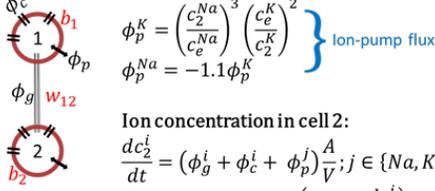

$\phi_p^K = \left(\frac{c_2^{Na}}{c_e^{Na}}\right)^3 \left(\frac{c_e^K}{c_2^K}\right)^2$ } Ion-pump flux
$\phi_p^{Na} = -1.1 \phi_p^K$

**Ion concentration in cell 2:**
$\frac{dc_2^i}{dt} = (\phi_g^i + \phi_c^i + \phi_p^j)\frac{A}{V}; j \in \{Na, K\}, \forall i \in \{Na, K\}$

$c_2^i(t+1) = c_2^i(t) + \left(\Delta t_{ion} \cdot \frac{dc_2^i}{dt}\right)$

**Membrane potential ($V_{mem}$) of cell 2:**
$v_2(t) = \frac{\sum_i c_2^i(t) z^i}{C} \frac{V}{A}, i \in \{Na, K, Cl\}$

**Signal flux with respect to cell 2 (nonlinear reaction-diffusion):**
$\widetilde{D_g}(t) = w_{12}.D_g(t).D_g$

$\phi_{c_2}^s(t) = \frac{1}{1 + e^{\widetilde{D_g}(t).(c_2^s(t) - c_1^s(t))}}$ {1st layer of Sigmoid}

$\widetilde{\phi_{c_2}^s}(t) = \frac{D_g(t).|w_{12}|.\phi_{c_2}^s(t)}{\sum |w_{12}|}$

$\frac{dc_2^s}{dt} = 2\left(\frac{1}{1 + e^{-20(\widetilde{\phi_{c_2}^s}(t) - b_2)}}\right) - 1$ {2nd layer of Sigmoid}

$c_2^s(t+1) = c_2^s(t) + \left(\Delta t_{sig} \cdot \frac{dc_2^s}{dt}\right)$

**Signal-gating of ion channels in cell 2:**
$D_c^{Na}(t) = c_2^s.\overline{D_c}$
$D_c^K(t) = (1 - c_2^s).\overline{D_c}$

**Voltage-gating of gap junction between cell 1 and cell 2:**
$D_{g_{c1}}(t) = \frac{1}{1 + e^{(-3.|w_{12}|.v_1(t))}}$

$D_{g_{c2}}(t) = \frac{1}{1 + e^{(-3.|w_{12}|.v_2(t))}}$

$D_g(t) = \frac{D_{g_{c1}}(t) + D_{g_{c2}}(t)}{2}$

**Glossary:**
$\phi_g^i$ – flux of ion $i$ across the gap junction
$D_g$ – diffusion coefficient of gap junction
$c_2^i$ – concentration of ion $i$ in cell 2
$c_e^{Na}$ – concentration of $Na$ ion in the environment
$\phi_c^i$ – flux of ion $i$ across the ion channel
$D_c^i$ – diffusion coefficient of ion channel for ion $i$
$\overline{D_c}$ – maximum diffusion constant of an ion channel
$D_g$ – base diffusion constant of a gap junction
$\overline{\phi_p^K}$ – flux of ion $K$ established by the ion pump
$\phi_g^s$ – flux of signaling chemical across the gap junction
$c_2^s$ – concentration of signaling chemical in cell 2 (AU)
$b_1, b_2$ – biases of cells 1 and 2
$w_{12}$ – meta-weight of the gap junction between cells 1 and 2
$v_2$ – $V_{mem}$ of cell 2
$A$ – surface area of a cell
$V$ – volume of a cell
$z^i$ – valence of ion $i$
$q$ – unit charge constant
$k_b$ – Boltzmann's constant
$T$ – Temperature
$F$ – Faraday's constant
$R$ – Gas constant
$C$ – Membrane capacitance
$\Delta t$ – time step size

**Constants:**
$z^{Na} = +1$
$z^K = +1$
$z^P = -1$
$c_e^{Na} = 145$ mol/m$^3$
$c_e^K = 5$ mol/m$^3$
$c_e^{Cl} = 150$ mol/m$^3$
$D_c^{Cl} = 0$ m$^2$/s
$g_{len} = 100e^{-9}$ m
$A = 4\pi r^2$ m$^2$; $r = 5.0e^{-6}$ m
$V = \left(\frac{4}{3}\right)\pi r^3$ m$^3$
$q = 1.602e^{-19}$ C
$k_b = 1.3806e^{-23}$ m$^2$ kg s$^{-2}$ K$^{-1}$
$T = 310$ K
$F = 96485$ C/mol
$C = 0.05$ F/m$^2$
$R = 8.314$ J/(K mol)
$\overline{D_c} = 150.0e^{-18}$ m$^2$/s
$\underline{D_g} = \frac{1.0e^{-18}}{g_{len}}$ m/s
$\Delta t_{ion} = 0.01$ (could vary)
$\Delta t_{sig} = 0.02$

**Initial conditions**
**(other than input cells):**
$c_j^{Na} = 10$ mol/m$^3$
$c_j^K = 125$ mol/m$^3$
$c_j^{Cl} = 135$ mol/m$^3$
$c_j^s = 0.5$ AU
$v_j = 0$ mV

$w_{12}, b_1, b_2$ are the learnable parameters

*Figure 8.* Mathematical details of the BEN model. (Top) The equations that describe the dynamics in a 2-cell system illustrated in the Fig. on the left. (Bottom) A glossary of variables used in the equations, the constants and initial conditions.

## Details of the simulation

We used the standard Euler integration method for integration the equations. We used a time step of 0.01 for the bioelectric dynamics and 0.02 for the signaling molecule dynamics. Each simulation was set up to run for a number of steps ranging from 300 to 600; smaller networks were run for fewer steps. Thus, the total simulation ranged between 300*0.01=3 simulation seconds to 600*0.01=6 simulation seconds.

**The backpropagation method for training elementary logic gates**

Regardless of the gate, a training run was set up as follows. A single BEN network was instantiated with randomized parameters. The initial weight parameters were chosen in the range [-1,1], and the biases in the range [0,1]. It was then fed with a batch of four inputs and was simulated for about 300 time-steps, with the inputs being held *fixed* throughout the simulation. An "input" constitutes a specific $V_{mem}$ pattern of the input cells. For example, input cell-1 may be set to a hyperpolarized $V_{mem}$ and cell-2 to a depolarized $V_{mem}$. The input cells were then randomly set to a different state, drawn from the training batch, without disturbing the state of the rest of the network (a method referred to as "continuous computation"[104]) and the simulation was continued for another 300 time-steps; this was repeated for ten times. At the end of each of 10 simulations, the observed states of the output cell during the last 10 time-steps were matched against the desired output corresponding to the input in the training batch (see Figs. 2b and 3b for examples). An error was then calculated based on the difference for each of the 10 simulations for each input sequence in the batch and averaged over. The derivatives of the final average error (gradient) with respect to all the parameters were then calculated. Due to the recurrent nature of the dynamics (Fig. 1), the simulation takes many more steps than the number of layers in the network before the output reaches an equilibrium. Thus, the gradients would in principle have to computed over all those steps—a method known as "backpropagation through time" (BPTT). However, due to computational constraints and known problems like "vanishing gradient", the gradients are typically computed over fewer time steps—this version of backpropagation is known as "truncated" BPTT. In this work, the gradients were computed over only the last 100 steps of a simulation. The quantum of changes for all the parameters were then calculated from the gradients after scaling it with a "learning rate" that set the pace of learning. The weights and bias were assigned separate "dynamic" learning rates, where the rates were fixed in such a way that the maximum change of a weight parameter was 0.1 and the maximum change of a bias parameter was 0.01. We adopted this strategy mainly to mitigate the vanishing gradient problem, and the assumption that small changes in the parameters should cause a smooth change in behavior. We also employed a "momentum" factor, a method often used in backpropagation, that determines the extent to which the previous parameter-change vector contributed to the following vector. We used a momentum of 0.5 for the first 100 training iterations, and 0.9 for the rest; the allowed range is [0,1]. Finally, the parameter-updates were applied (thus error was backpropagated), and the whole process was repeated for about 600,000 "epochs". The weights were allowed to change infinitely in both directions, whereas the biases were limited within the range [0,1]. We deemed a network as successfully trained if it achieved an average performance error of 0.0002 or below at some point during the training.

We used the Python software package called *Pytorch* for computing derivatives during backpropagation. This package uses a method called "automatic differentiation" to compute derivatives based on "computational graphs"—a graph that keeps track of every computation performed during a simulation, which is then swept backwards for computing the derivatives, essentially constituting an algorithm for the "chain rule" of differential calculus.

**The combined backpropagation-genetic-algorithm method for training tissue-level logic gate and the pattern detector**

A population of 50 BEN "genomes" was instantiated. A genome consists of a combination of the network's parameters, namely the weight matrix and the bias vector, that define a network. Henceforth, by "genome" we refer to a BEN network, for clarity. Each network consists of the following structure: an input layer consisting of 2 nodes, an intermediate "tissue" layer consisting of 25 nodes, and the output layer consisting of a single node. Initially the tissue layer was designed to be a two-dimensional lattice. Next, the tissue layer of each individual in the population was randomly rewired with a rewiring probability chosen uniformly from the range [0,1]. Thus, the population consisted of the full spectrum of tissue layer structures ranging from lattice to random. Next, the input layer and the output layers of each individual in the population were respectively connected to approximately 50% of the nodes in the tissue layer; this gave the GA a chance to explore both denser and sparser inter-layer connectivity. Every individual in the population was then assigned randomized weight and bias parameters chosen in the same range as backpropagation. They were each then backpropagated for 80 iterations in the same fashion as described above for the elementary gates, with an important difference: the inputs for these gates were *transient*, as they are set for the first time-step only. The final average error of each individual became its fitness score. The GA then goes through the conventional steps of selection, mutation, creating a new generation of individuals.

We used a simple flavor of GA known as the "microbial genetic algorithm"[136] that capitalizes on the notion of horizontal gene transfer observed in microbes. It is essentially a form of tournament selection where two individuals in a population are picked and pitted against each other (their fitness scores are compared). We used a "geographical selection" method where individuals are placed on a 1D ring, and only geographically close individuals are picked for the tournament. The size of selection-neighborhood is known as the "deme size", which was set to 10% in this work. The genome of the winner is transmitted to the loser at a certain "recombination rate", which was set to 0.1. The genome of the loser is then mutated at a certain "mutation rate" (set to 0.05) and reinserted into the population. This process is repeated for 1000 generations. The individual with the best fitness score in the final generation was deemed as the best individual—the most successfully trained tissue-level logic gate.

**Generation of the input patterns during the training of the pattern detector**

The set of input patterns used for training the pattern detector consists of two categories: (1) French-flag (FF) and (2) Not-French-flag (NFF). The subset FF consists of the French-flag pattern itself and its slightly noisy variants. The latter is generated by perturbing each pixel of the French-flag approximately within the range [-24mV, +24mV]. Any perturbation that takes a pixel below the bounds of [-80mV, +80mV] are reset to the closest boundary values. The subset NFF consists of the more distorted variants of the French-flag. It is generated in the same way as FF except the perturbation range is [-40mV, +40mV]. A total of 20 samples in each set were generated during each iteration of the GA-BP search.

## Acknowledgements

We thank Alexis Pietak and many members of the Levin lab for many helpful discussions, as well as Nima Dehghani and Joshua Finkelstein for helpful comments on a draft of the manuscript. This research was supported by the Allen Discovery Center program through The Paul G. Allen Frontiers Group (12171), and M. L. gratefully acknowledges support from the Templeton World Charity Foundation (TWCF0089/AB55), the Barton Family Foundation, and the National Science Foundation (DBI-1152279 and Emergent Behaviors of Integrated Cellular Systems subaward CBET-0939511). This research was also sponsored by the Defense Advanced Research Projects Agency (DARPA) under Cooperative Agreement Number HR0011-18-2-0022, the Lifelong Learning Machines program from DARPA/MTO. The content of the information does not necessarily reflect the position or the policy of the Government, and no official endorsement should be inferred. Approved for public release; distribution is unlimited.

**Data Availability Statement**

All parameters used for the networks in this manuscript, as well as the source code, can be downloaded at https://gitlab.com/smanicka/BEN .

# "Modeling somatic computation with non-neural bioelectric networks"

Santosh Manicka and Michael Levin*

Allen Discovery Center
200 College Ave.
Tufts University
Medford, MA 02155

* Author for correspondence:
    Email: michael.levin@tufts.edu
    Tel.: (617) 627-6161

# Supplementary information

## 1. Parameters of trained BEN models

All parameters used for the networks in this manuscript can be downloaded at https://gitlab.com/smanicka/BEN. The parameters of a BEN network that are learned during training, regardless of the problem they are trained to solve, are the 'weights' and 'bias' (highlighted in red in Fig. 8). The weights represent the strengths of the connections between cells (hence a matrix), determining the direction and speed by which the corresponding gap junctions alter their permeabilities dynamically as a function of $V_{mem}$. The biases represent the thresholds of the cells (hence a vector), determining how the signaling molecule switches speed and direction of change in concentration in the cells. The data files contain the weight and bias details of the best network from each of the five experiments described in the main text (simple AND logic gate, simple XOR logic gate, tissue-level AND, pattern detector and the compound NAND logic gate). File names are self-explanatory. Each file contains labels like 'Cell 1', 'Cell 2' etc. indicating the cell numbers. In the case of the compound NAND logic gate, the labels also indicate which module a cell belongs to (AND, BRIDGE or NOT), for example, 'AND Cell 1', 'BRIDGE Cell 10', 'NOT Cell 14' etc. The ordering convention adopted for numbering the cells is illustrated in Supplementary Fig. S1.

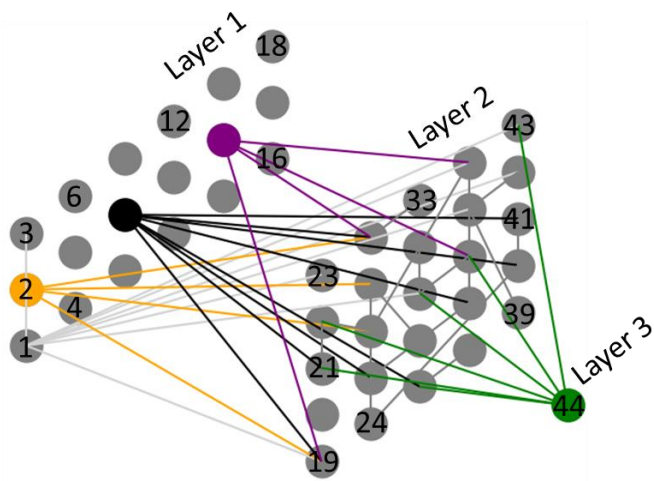

*Supplementary figure S1.* Numbering convention of the cells in a BEN network. Shown are a few examples of cell numbers in a sample network; the rest can be inferred. The lower left cell of the first layer is always numbered 1, while the highest numbered cell is the top right cell in the last layer.

## 2. BEN achieves robustness to damage by distributing information

Here, we present an analysis of the robustness of a nine-cell AND gate using the tools of information theory[125]. We wanted our network to be large enough to contain robustness characteristics (due to redundancy of function among the cells) but small enough to be amenable to analysis. First, we trained a nine-cell AND logic gate using backpropagation as described in the main text. Next, we knocked-off individual nodes that were neither input nor output cells, and then measured the performance of the resulting networks. Finally, we correlated the net amount of information about the inputs and outputs flowing through the rest of the nodes with the network performance upon removal of those nodes. We computed the information flow in the network using a Python package 'IDTxl'[125] that computes the conditional transfer entropy (the amount of mutual information between a source's past state and a target's present state, given the target's past states) for every pair of nodes in the network (Supplementary Fig. S2).

We conclude that the network is more robust to removal of nodes through which less information flows (Supplementary Fig. S3). Moreover, there is modularity in the network, where some cells contain more information about one input versus the other (Supplementary Fig. S4). For example, cell 7 contains disproportionately more information about input 2 than input 1, while cells 2 through 6 contain slightly more information about input 1 compared to input 2. The biological implication is that real somatic tissues may also possess such robustness of function to damage of cells, and the mechanism by which this occurs may be distributed information processing, redundancy and modularity.

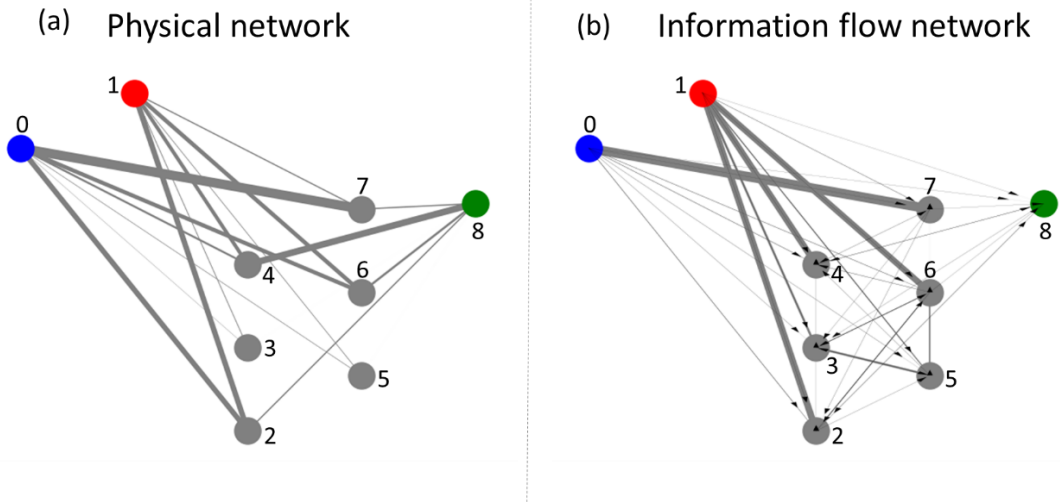

*Supplementary figure S2*. Physical and information-flow network depictions of a 9-cell AND gate. Color coding follows the same convention as in the main text. The edge weights of the physical network represent "meta weights" (see main text). The edges in the information-flow network are both weighted and directed: weights represent the amount of information-flow (measured by conditional transfer entropy) and directions represent the direction of the flow. Notice that corresponding weights are similar in the two networks, although there are some noticeable differences as well. Lastly, the information-flow network contains edges that are not present in the physical network; for every such edge an indirect path can be traced in the physical network.

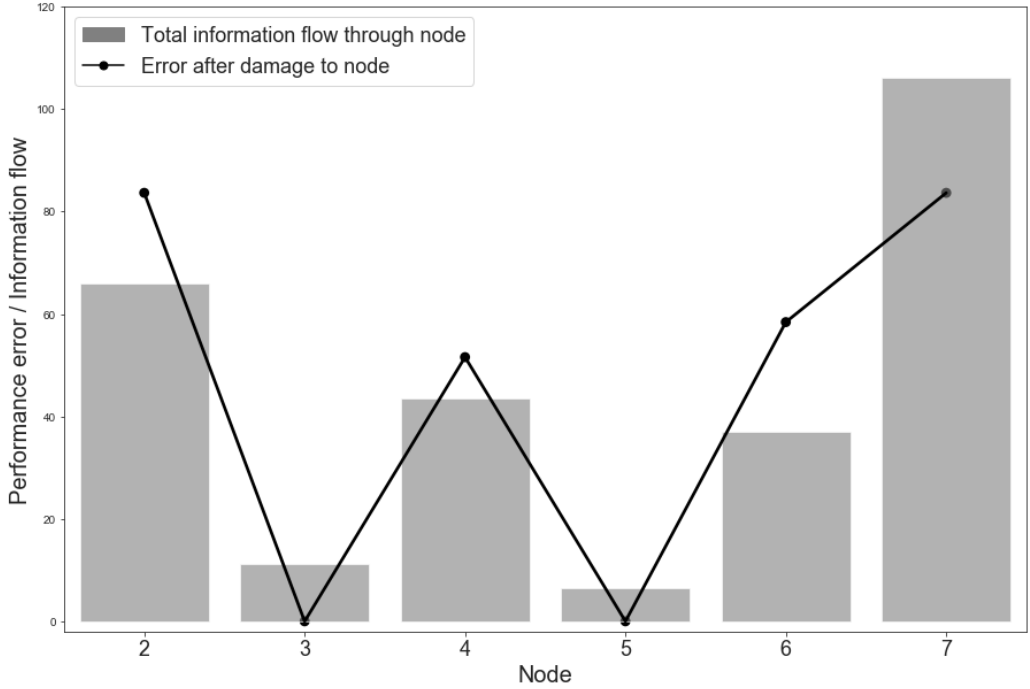

*Supplementary figure S3.* The relationship between information-flow and robustness to node damage. The heights of the bars represent the total amount of information about the inputs and output combined flowing through a node. Clearly, nodes that conduct more information are relatively more important to the functioning of the network: disabling them results in higher performance error of the whole network.

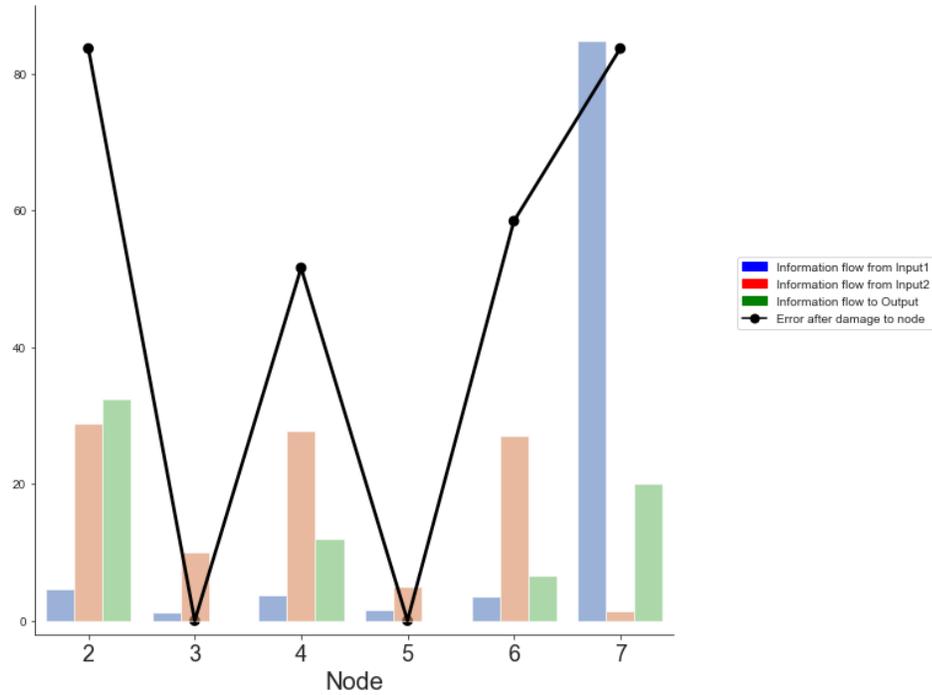

*Supplementary figure S4.* Modularity in information-flow. A form of specialization has emerged in this network: node 7 conducts disproportionately more information about input 1 compared to input 2 than any other intermediate cell, while all other cells conduct relatively more (but not disproportionately more) information about input 2 than about input 1. This segregation of knowledge about the inputs reflects the generic notion of modularity that biological systems are replete with both in terms of structure and function.

### 3. Alternative approaches to training BEN models

BPTT is just one of the many possible machine learning methods that can be used to train BEN models. To demonstrate that BEN networks can in principle be trained using other methods, we chose an alternative to gradient descent namely evolutionary optimization; genetic algorithms (GA) is an example of this class. Here, we used a GA to train a BEN network to function as an AND gate. Note that we didn't combine BP with GA for this example, unlike the tissue-level AND gates and the pattern detector discussed in the main text. We used the microbial GA (see 'Methods' section of the main text) with a population of 100 BEN networks, and a range of parameters including a recombination rate of 7-10%, a mutation rate of 5-10% and a deme size of 5-10. We found that none of the runs evolved a network to attain the optimum error (Supplementary Fig. S5), demonstrating the inferiority of GA compared to BPTT for training BEN networks. We then implemented a run by seeding the population with a previously best-known AND gate obtained through BPTT training, and injected Gaussian noise into the population. This led to the evolution of a BEN network that crossed the optimum error threshold (black line in Supplementary Fig. S5).

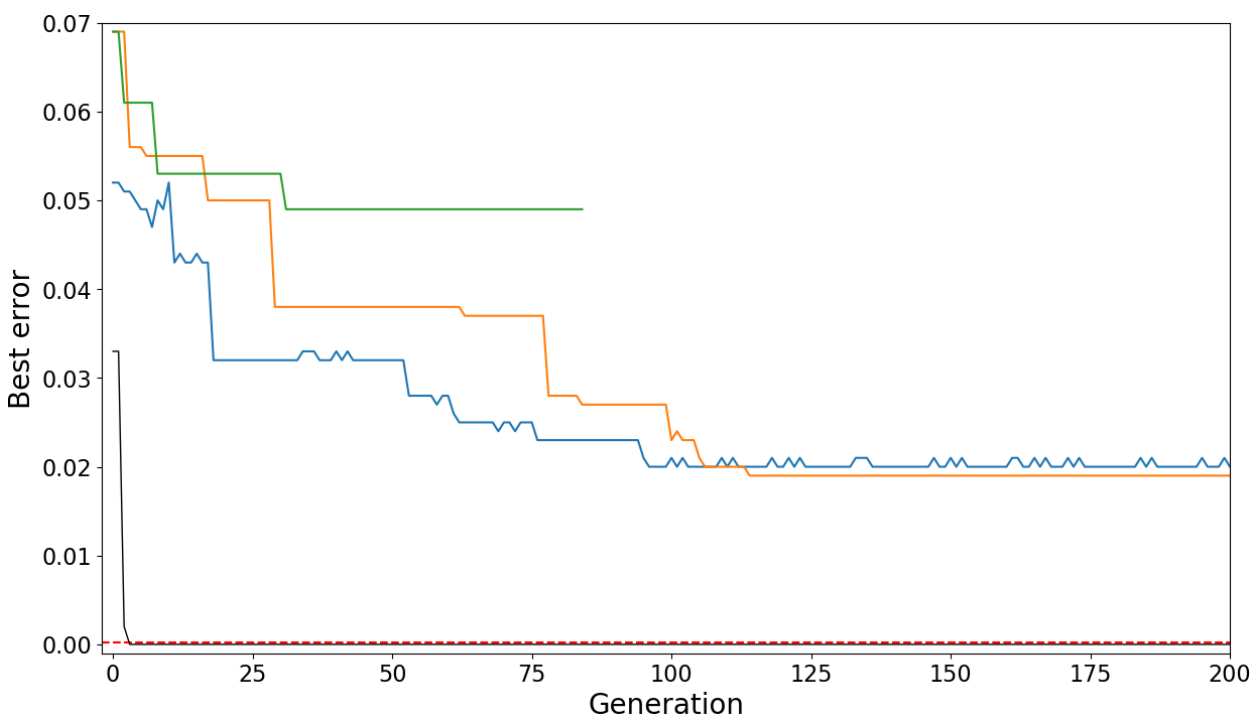

*Supplementary figure S5.* Performance errors of the best individuals in each iteration of the GA from four different runs indicated by the different colors. The black line represents a run with a population that was seeded with a previously trained (with BPTT) AND gate; Gaussian noise was injected into the population before initiating the run. The horizontal red dashed line indicates the optimum error required for a BEN network to qualify as a logic gate.

### 4. Designing and training a compound logic gate

Constructing a compound logic gate using pre-trained modules requires more than simply connecting the two modules. This is because, unlike genetic networks and neural networks which are directed, BEN networks are bidirectional due to the symmetric nature of gap junctions. Therefore, if the modules are simply connected together, then the output of one module will also receive signals from the input of the connected module. This could potentially result in the upstream module outputting the wrong state, and thus the downstream module receiving and outputting the wrong states as well. To mitigate it, we included an "asymmetric bridge" layer (Supplementary Fig. S6) that interfaces the modules and trained it so that the compound gate as a whole behaves as desired: the downstream module outputs the correct state for every set of inputs received by the upstream module. The asymmetric aspect of the bridge lies in the differential connectivity to the upstream and downstream modules. First, only one node in the bridge is connected to the output of the upstream module, thus effectively creating a bottleneck between the two modules and minimizing the influence on the upstream output due to downstream noise. Second, all of the rest of the bridge nodes are connected to the input of the downstream modules=, thus maximizing the chances of passing down information reliably from the upstream module. For this particular experiment, we chose a bridge consisting of a single layer of 10 nodes connected as a chain. We explored various settings for training the NAND gate: (1) Specified the states of the inputs and outputs as in any other logic gate; (2) specified the states of the output and the input nodes of the upstream and downstream modules respectively; (3) trained the AND and NOT modules by accommodating intrinsic activity, as in the French-flag detector; and (4) trained the AND and NOT modules by outputting a $V_{mem}$ of 0 mV when inputs are equal to 0 mV, thus creating a "scaffold" attractor for the modules to use as an intermediate step to overcome the effects of inter-module dynamics. We found that settings (3) and (4) were crucial to training the NAND gate. In particular, setting (4) manifests itself during the last phase of the simulation when both inputs are high; the output lingers around 0 mV (the scaffold attractor) for a while before "hammer-throwing" itself out of it into the correct attractor (time point 4 in Fig. 7).

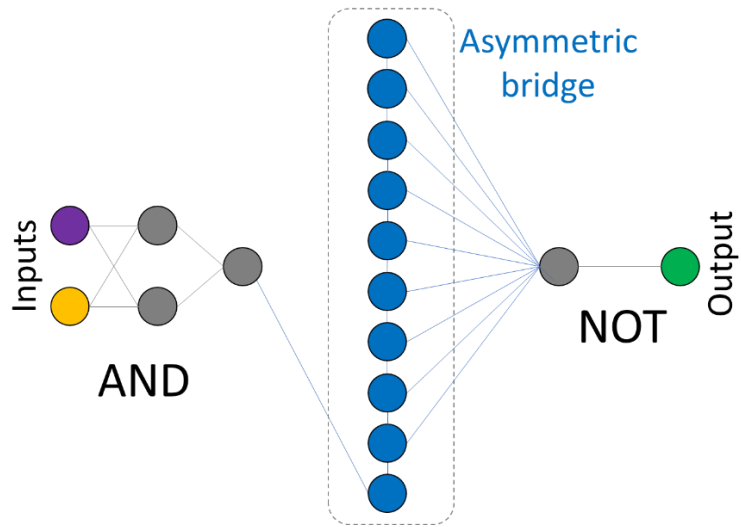

*Supplementary figure S6*. The architecture of a compound logic gate, specifically a NAND gate. The overall layout consists of three parts, namely the upstream (AND), the bridge and the downstream (NOT) modules. The AND and NOT modules are pre-trained, and only the bridge (blue) is trained here. The overall gate still consists of two inputs (orange and purple nodes) and a single output (green).